# Strain tunable semimetal-topological insulator transition in monolayer 1T'-WTe$_2$

Chenxiao Zhao[a#], Mengli Hu[b#], Jin Qin[a], Bing Xia[a], Canhua Liu[a,c,d], Shiyong Wang[a,c,d], Dandan Guan[a], Yaoyi Li[a,c,d], Hao Zheng[a,c,d], Junwei Liu[*b], Jinfeng Jia[*a,c,d]

[a] *Key Laboratory of Artificial Structures and Quantum Control (Ministry of Education), Shenyang National Laboratory for Materials Science, School of Physics and Astronomy, Shanghai Jiao Tong University, Shanghai 200240, China*

[b] *Department of Physics, Hong Kong University of Science and Technology, Hong Kong, China*

[c] *Tsung-Dao Lee Institute, Shanghai Jiao Tong University, Shanghai 200240, China*

[d] *CAS Center for Excellence in Topological Quantum Computation, University of Chinese Academy of Sciences, Beijing 100190, China*

[#] C. Zhao and M. Hu These authors contributed equally to the work.

liuj@ust.hk; jfjia@sjtu.edu.cn

**Abstract:**

A quantum spin hall insulator (QSHI) is manifested by its conducting edge channels that originate from the nontrivial topology of the insulating bulk states. Monolayer 1T'-WTe$_2$ exhibits this quantized edge conductance in transport measurements, but because of its semimetallic nature, the coherence length is restricted to around 100 nm. To overcome this restriction, we propose a strain engineering technique to tune the electronic structure, where either a compressive strain along *a* axis or a tensile strain along *b* axis can drive 1T'-WTe$_2$ into an full gap insulating phase. A combined study of molecular beam epitaxy and in-situ scanning tunneling microscopy/spectroscopy then confirmed such a phase transition. Meanwhile, the topological edge states were found to be very robust in the presence of strain.

Monolayer transition metal dichalcogenides (TMDs) in 1T' structure are predicted to be QSHIs [1]. Within this family, WTe$_2$ and WSe$_2$ have been verified experimentally using angle-resolved photoemission spectroscopy (ARPES) [2-4] and scanning tunneling microscopy/spectroscopy (STM/STS) [2-5]. Moreover, monolayer 1T'-WTe$_2$ has already been confirmed to be topological nontrivial by various nonlocal transport measurements [6] and exhibits a fully quantized conductance up to 100 K [7]. However, the coherent edge channel length is found to be around 100 nm because of the semimetallic nature of the bulk states [7], hindering both further fundamental research and potential applications. To overcome this hindrance, it is necessary to manipulate the electronic structure of monolayer WTe$_2$ to realize a full gap insulator. In principle, the electronic structure of a material can be engineered using various external fields such as electric or mechanical strain fields. Recently, electrical tuning has been applied extensively to TMD materials for tuning structures [8], superconductivity [9, 10] and Berry curvature dipole [11]. For low-dimensional systems, strain engineering is a very general and effective approach and is technologically feasible. Examples include strain-induced topological phase transition [12], strain engineering Dirac states [13, 14], strain-induced flat band [15, 16], strain tunable magnetism at domain walls [17], strain-induced surface-dominated conduction [18] and strain-enhanced superconductivity [19]. However, related studies in TMD family are relatively limited and unproved [1, 20, 21]. So far, no method has been found to tune the band structures of monolayer 1T'-WTe$_2$ effectively.

In this work, based on orbital analysis and the first-principles calculations, we proposed to use strain to optimize the band structures of monolayer 1T'-WTe$_2$, and we found that a compressive strain along *a* axis or a tensile strain along *b* axis can convert monolayer WTe$_2$ from a semimetal to an insulator. This was further confirmed by a systematic study using molecular beam epitaxy (MBE) and STM, where strains were introduced by the distorted substrate. The strain status of WTe$_2$ films was measured by STM, and the corresponding band structures were characterized by differential conductance (dI/dV) spectra. Based on experimental data, we find that a moderate in-plane strain can open a full insulating gap in WTe$_2$, and the gap size is positively related with lattice constant *b* and negatively related with *a* (*a* and *b* are marked in Fig. 1(a)), consistently with the theoretical results. Moreover, the topological edge states are obtained on both semimetallic and insulating WTe$_2$ islands, indicating that the nontrivial topology is robust in the presence of external strains. Our results unambiguously demonstrate the feasibility of strain engineering to realize the fully gapped QSHI in monolayer 1T'-WTe$_2$, making 1T'-WTe$_2$ a suitable material platform for developing controllable topological electronic devices with a long coherence length.

Monolayer 1T'-WTe$_2$ has a rectangular unit cell as shown in Fig. 1(a), where one W-atom layer is sandwiched between two Te-atom layers. Moreover, there is a Peierls distortion along the *a* axis, inducing a large band inversion at Γ point and driving monolayer 1T'-WTe$_2$ into a Z$_2$ topological phase [1]. However, a finite overlap between the valence band and conduction band renders it semimetallic [1, 22]. To determine a suitable approach to manipulate the band structures, we first analyzed the orbital compositions of states at Γ point around the Fermi level. As shown in Fig. 1(b), the states of the conduction and valence bands (marked out by gray dashed boxes) are dominated by $d_{xz}$ orbitals from W atoms and $p_y$ orbitals from Te atoms, respectively (contributions from various orbitals are shown in Fig. S1). The spatial distributions of these orbital wave functions are plotted schematically in Fig. 1(c). Along *a* direction (same with $x$ axis), the $d_{xz}$ orbitals form $dd\pi^*$-like anti-bonding states, while $p_y$ orbitals form $pp\pi$-like bonding states. Upon applying a tensile strain, both the $dd\pi^*$-like and $pp\pi$-like bonds are weakened, resulting in the energy of conduction band decreasing and the energy of valence band increasing, thus enhancing their overlap. On the other hand, a compressive strain along the $a$ direction leads to the opposite energy shifts, thus reducing and even erasing the band overlap. Moreover, both $d_{xz}$ and $p_y$ orbitals are anisotropic and more specifically their responses to the mirror flip along $a$ and $b$ directions are opposite, which results in opposite responses to the strain along the $b$ (same with $y$ axis) direction. In details, the $d_{xz}/p_y$ orbital forms $dd\delta$ / $pp\sigma^*$ -like bonds, leading to an energy increase or decrease in the conduction/valence band with tensile strain and an opposite change with compressive strain. Thus, tensile strain along the $b$ direction can also remove the band overlap. The strain effects along the $z$ direction were also studied and shown in supplementary material II [23].

The responses to the strain from the orbital analysis above is confirmed by the first-principles calculations (more details in supplementary material III [23]). The fundamental gap values (calculated using the HSE03 hybrid functional [24, 25]) as a function of strains along the $a$ and $b$ directions are shown in Fig. 1(d) and 1(e), respectively. It is clear that the monolayer 1T'-WTe$_2$ tends to open a fundamental gap under a compressive (tensile) strain along $a$ ($b$) direction. Calculations using different exchange-correlation functionals, including generalized-gradient approximation [26] and local-density approximation [27], show similar strain responses, as shown in Fig. S5. All the results indicate that the band structures of monolayer 1T'-WTe$_2$ can be manipulated effectively by strain to realize a full gap insulator.

To confirm the theoretical finding, we fabricated strained monolayer 1T'-WTe$_2$ using MBE

and systematically investigated the strain effects using STM/STS. After many attempts, we found the strain can be induced by using a distorted graphene substrate treated by over-heating 6H-SiC (see experimental details in methods [23]). The topography and atomic resolved image in a flat region (see Fig. 2(a)) confirm the quality of the WTe$_2$ films. The measured lattice constants are $a$ = 6.45±0.03 Å and $b$ = 3.46±0.03 Å, respectively. The dI/dV spectra taken from different flat regions are shown in Fig. 2(c), which show uniform semimetallic characters. The minimum of the local density of states (LDOS) is ~ 60 meV below the Fermi level (manifested as a dip) and a V-shape gap locates at zero bias, consistent with previous STM results [5, 22]. This V-shape gap is the Coulomb gap resulting from the electron-electron interactions in monolayer WTe$_2$, which always exists while the LDOS at the Fermi level is nonzero [5, 22]. When the substrate is deformed, the crystal structure of WTe$_2$ films grown on it will also be distorted. Figure 2(b) gives an example of the distorted regions, with the WTe$_2$ films showing clear deformations. Three typical small distorted areas with different lattice constants, i.e., small parts of several nanometers taken from the large films, are shown in Fig. 2(d)-2(f) (top panels) together with the dI/dV spectra (bottom panels) taken from the center of each area. Different from the flat regions, the dI/dV spectra of distorted regions are non-uniform and may exhibit either semimetallic behavior with a dip and a V-shape coulomb gap (Fig. 2(d)) or insulating behavior with a large U-shape gap (Fig. 2(e) and 2(f)). In addition, the size of the U-shape insulating gap in different areas also differs (also see Fig. S6 (a)). From the spatially resolved spectra in Fig. S3, S6 (a) and S7, one can see that the U-shape gap exhibits clear strain-dependence, especially in Fig. S3, where the U-shape gap starts from a dip below the Fermi level and gradually becomes a full gap with zero LDOS when the strain is large enough. During this process, the V-shape coulomb gap is initially unchanged and eventually disappears when the LDOS around Fermi level becomes zero because of the increased U-shape gap. In addition, we also doped the WTe$_2$ films with potassium (K) atoms to shift the Fermi level to exclude the effect of Coulomb interactions on the gap. As shown in the Fig. 2(g), the Fermi level is upper shifted ~40 meV and the dip now resides around -100 meV. At some locations in distorted regions, the strain induced U-shape gap still holds even though the Fermi level is far from it. Thus, we can safely rule out the possibility that the insulating gaps are caused by coulomb blockage or other types of many-body effect. The influences of substrates are also excluded in supplementary material V [23]. Combining all the evidence above, it is reasonable to deduce that the insulating gap is induced by strain, and we further performed a more detailed study of gap size and lattice constants.

We took dozens of sets of experimental data in different areas of distorted regions, and each set includes a gap value and the corresponding in-plane lattice constants (*a*, *b*). As shown in Fig. 3(a), the gap decreases with *a* but increases with *b*, and is zero when *a* is large and *b* is small. To study uniaxial dependency of gap on lattice constants *a* and *b* quantitatively, we regrouped the data and averaged the gaps with the similar *a* or *b* in Fig. 3(b) and (c). We examined the analysis very carefully to rule out any artefact caused by different choices of averaging methods (see more details in supplementary material VII [23]). We found that the gap will decrease with *a* over the range from 6.23 Å to 6.54 Å, but increase with *b* over the range from 3.49 Å to 3.57 Å. In addition, the gap closes when *a* is larger than 6.54 Å (~ 1.40% tensile strain) or *b* is smaller than 3.49 Å (~ 0.87% compressive strain). These experimental results are consistent with our theoretical predictions. Besides, the strain effects along the *d* axis were also qualitatively analyzed in these experiments, as shown in supplementary material II [23].

All the theoretical and experimental results above clearly demonstrate that in-plane strain can indeed be used as an effective tool to tune the band structures of monolayer 1T'-WTe$_2$ and drive a phase transition from a semimetal to an insulator. Now the critical question is whether the nontrivial $Z_2$ topology still holds under strain. To answer this question, we first analyzed the phase diagram for strains between -10% and +10%, as shown in Fig. 4(a). The whole phase diagram is divided into three different regions: insulating nontrivial phase, semimetallic nontrivial phase and semimetallic trivial phase. The first two phases dominate the phase diagram, indicating the persistence of nontrivial topology subject to external strain fields, which is due to the large band inversion at Γ point induced by Peierls distortion. All experimental data points are located within these two nontrivial regimes and are represented by black circles in the phase diagram. Moreover, the freestanding WTe$_2$ without strain is near the phase boundary of the insulating nontrivial phase and semimetallic nontrivial phase, which makes the strain a feasible and efficient approach to phase engineering.

In experiments, the gapless states located at edges are strong evidences for QSHI. In flat regions, we obtain edge states similar to previous studies [5] (see Fig. S10), confirming the nontrivial $Z_2$ topology. In distorted regions, we focus on insulating WTe$_2$ islands with a full bulk gap, such as the one shown in Fig. 4(b) (see more examples in Fig. S11). Typical dI/dV spectra taken on it are shown in Fig. 4(c), where a full gap as large as ~ 50 mV is observed in the bulk while metallic spectra are obtained at two edges along both the *a* and *b* directions, labeled as A and B edges. It is worth noting that the V-shape coulomb gap at zero bias again appears in the edge states since there are finite LDOS at the Fermi level from the topological edge states, even when there is a large insulating U-shape gap in the bulk states. Using the experimental lattice constants of this WTe$_2$ island, we calculate the band structure along the *a* direction (in Fig. 4(d)). There are indeed topological edge states spanning the whole bulk gap, which is consistent with the metallic dI/dV spectrum taken at the A edge. Similar results for the B edge are shown in Fig. S12.

Besides confirming the nontrivial topology, we further studied the penetration length of the topological edge states, which is particularly important when designing devices. Two series of dI/dV spectra were measured along the dotted lines in Fig. 4(b), as shown in Fig. 4(e) and 4(f). Clearly, the gap vanished near the edges and the intensity of in-gap states increases rapidly indicating the existence of topological edge states. To obtain the concrete value of the penetration length, the spatial distributions of integrated intensity of edge states from -20 mV to -50 mV are shown at the bottom panel, indicating a penetration of ~ 3.0 nm for the A-edge state and ~ 3.6 nm for the B-edge state, respectively. Such a short penetration length implies that the quantized conductance can be confined to a narrow channel, which is vital for realizing high-density dissipationless topological devices.

In summary, strain engineering of the band structure of 1T'-WTe$_2$ was theoretically predicted and experimentally confirmed. We found that strain can tune the electronic properties and induce a phase transition from a semimetal to an insulator, while maintaining the nontrivial $Z_2$ topology. The tunable band structure and robust topology make WTe$_2$ a suitable material platform for both fundamental research and topological electronic devices.

**Figures:**

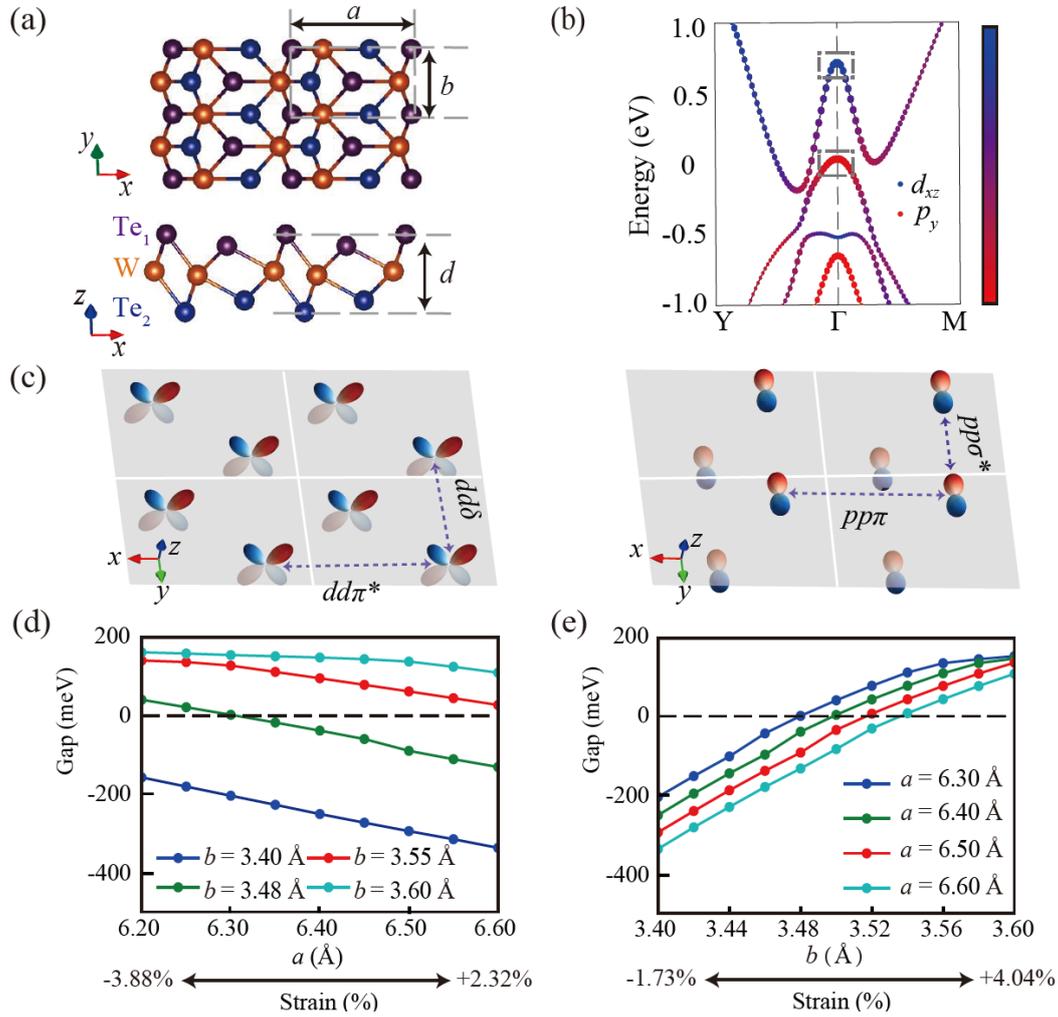

Fig. 1. Theoretical analysis of strain effects in monolayer 1T'-WTe$_2$. (a) Crystal structure of monolayer 1T'-WTe$_2$. Top view (upper panel) and side view (lower panel). Three atomic layers are denoted by Te$_1$, W and Te$_2$. The in-plane lattice constants $a$ (along $x$ direction) and $b$ (along $y$ direction) and the distance between two Te atomic layers $d$ (along $z$ direction) are also marked. (b) Orbital compositions for states around the Fermi level. (c) Schematic plot of wave-functions of $d_{xz}$ (left panel) and $p_y$ orbital (right panel), respectively. Each panel contains 4 unit cells, and the bonding conditions between unit cells are indicated by double sided arrows. (d, e) Gap evolution as a function of lattice constants $a$ and $b$, respectively, calculated by using the HSE03 hybrid functional.

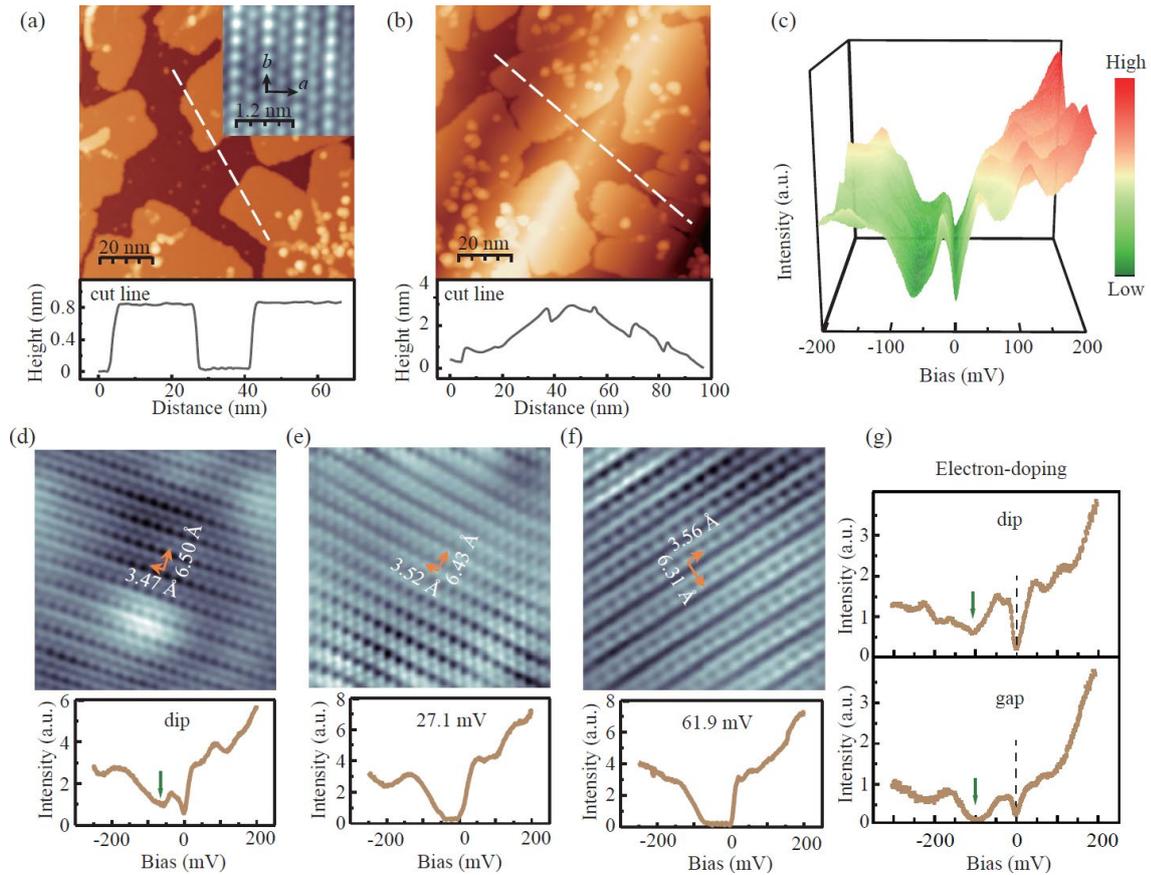

Fig. 2. Monolayer 1T'-WTe$_2$ films with and without strain. (a) and (b), Topography of 1T'-WTe$_2$ in flat and distorted regions, respectively (U = 2.0 V, I$_t$ = 100 pA, T ~ 4.2 K). Profiles along the white dashed lines are shown at the bottom. Inset: atomic resolved image taken in a flat region (U = 45 meV, I$_t$ = 1 nA, T ~ 4.2 K). The *a* and *b* axes are marked by black arrows. (c) Typical dI/dV spectra taken in different areas of flat regions. (d-f), Top panels: typical atom-resolved images in deformed regions (d: 7 nm×7 nm, U = 20 mV, I$_t$ = 1 nA; e: 7 nm ×7 nm, U = 20 mV, I$_t$ = 3.7 nA; f: 7 nm ×7 nm, U = 50 mV, I$_t$ = 2.45 nA. T~ 4.2 K). Bottom panels: the corresponding dI/dV spectra taken in the center. (g) The dI/dV spectra taken after tuning the Fermi-level by electron-doping using potassium (K) atoms. Top panel: spectra with a dip taken in the flat region; Bottom panel: spectra with a gap taken in the distorted region.

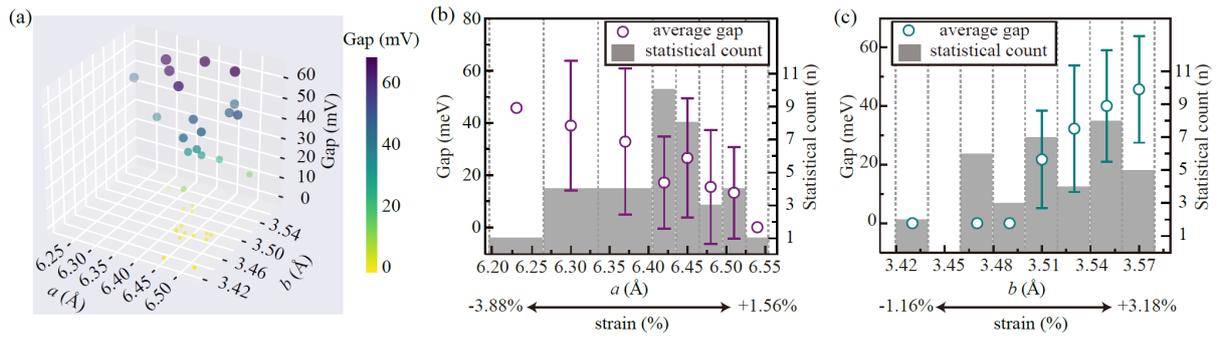

Fig. 3. Experimental relations between average gap values and lattice constants. (a) 3D view of gap value as a function of in-plane lattice constants $a$ and $b$; (b, c) Dependency of gap on strains along the $a$ or $b$ directions. Each data point in (b) and (c) is an averaged gap value within a lattice interval, divided by gray vertical dashed lines, and the standard deviations of gap values are represented by the vertical bars. The histograms show the statistical count in each lattice interval, corresponding to the right vertical ordinates. The converted strain values based on lattice constants in flat regions are also shown at the bottom.

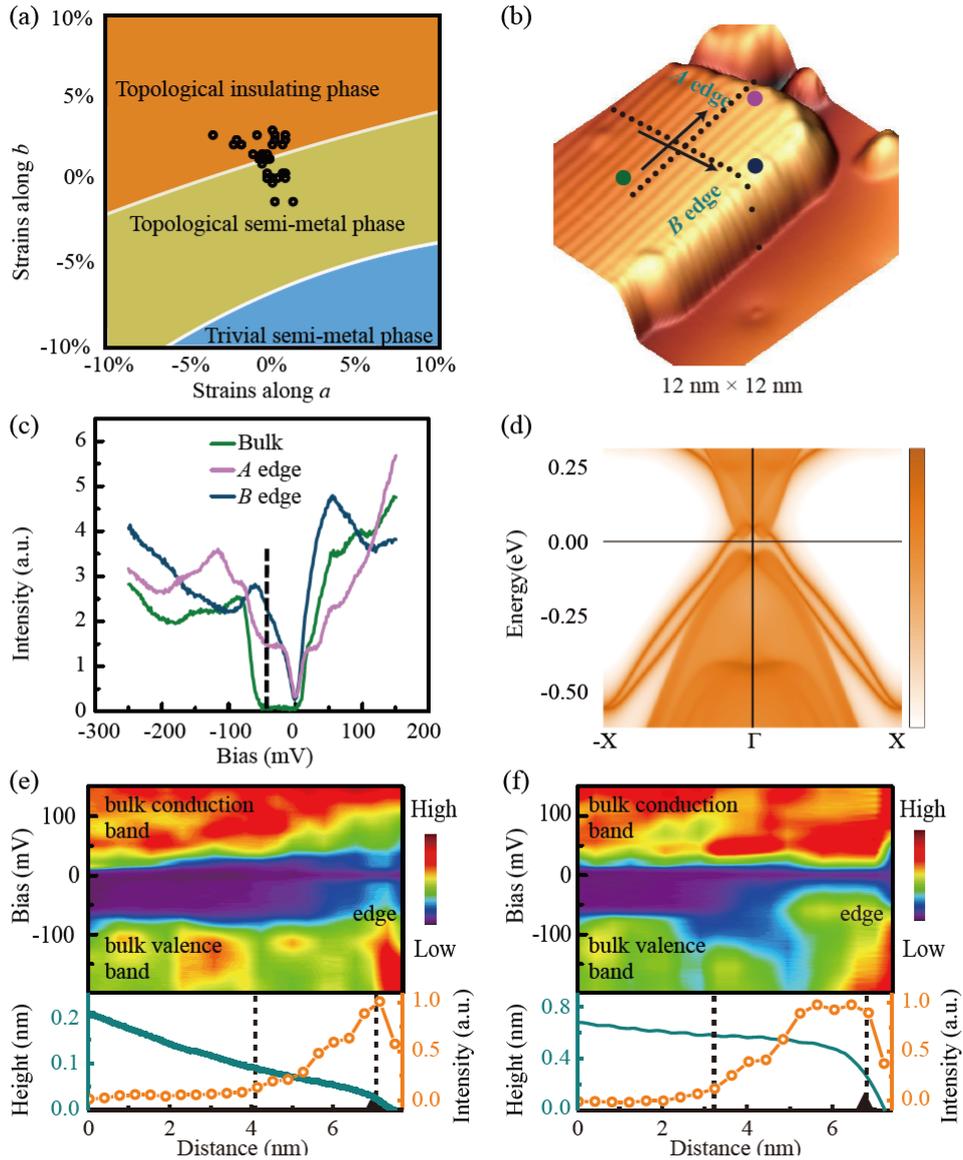

Fig. 4. Topological properties of the insulating 1T'-WTe$_2$. (a) Phase diagram of 1T'-WTe$_2$ as a function of in-plane strain. Strain conditions of experimental data are marked by black circles. (b) Topography of a small WTe$_2$ island with a full bulk gap in the distorted region. Two edges along the *a* and *b* axes are denoted by A and B, respectively. (c) Typical dI/dV spectra taken in the bulk and at two different edges. The corresponding locations are marked by points with different colors in (b). (d) Calculated edge states along the A edge. The lattice constants used here are experimental values measured on the island: 6.33 Å and 3.54 Å. (e, f) Spatially resolved color images of a series of dI/dV spectra taken crossing the A edge and B edge,

respectively (along dotted lines in (b)). Spatial distributions of the integrated intensity of edge states from -20 mV to -50 mV are extracted from the color images and are shown at the bottom by orange circles. The penetration lengths of edge states into the bulk are represented by the distance between two vertical dotted lines. The height profiles are shown by cyan curves.

# Supplementary materials

# I. Orbital projections around Fermi level

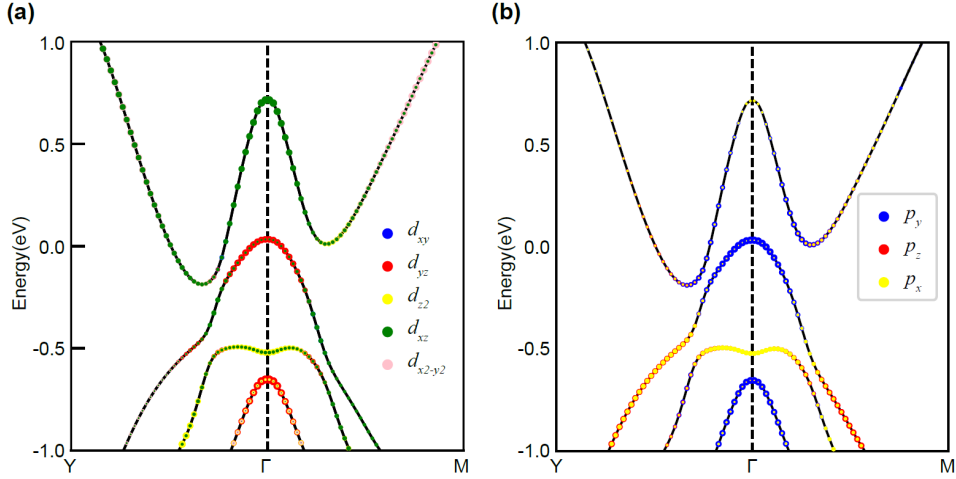

Fig. S1. Contributions from different orbitals. (a) Orbital projections from W atoms. (b) Orbital projections from Te atoms. Marker size represents the contributions for specific orbital.

Fig. S1 shows the orbital projections from W and Te atoms, respectively. A larger marker means more contributions by the orbital. According to the results, the dominant orbital compositions of conduction band (CB) and valence band (VB) at $\Gamma$ point around Fermi level are $d_{xz}$ and $p_y$, respectively. It is noted that $d_{yz}$ orbital also contributes a lot for VB as shown in Fig. S1(a). But the participation of $d_{yz}$ orbital will not alter our analysis in the main text. Because that the responses of $d_{yz}$ and $p_y$ orbitals to mirror flip along *a* and *b* directions are identical, thus they have the same effects when applying a strain.

## II. Strain effects along $z$ direction

### 1. Orbital analysis and first-principles calculations of Z-distortions

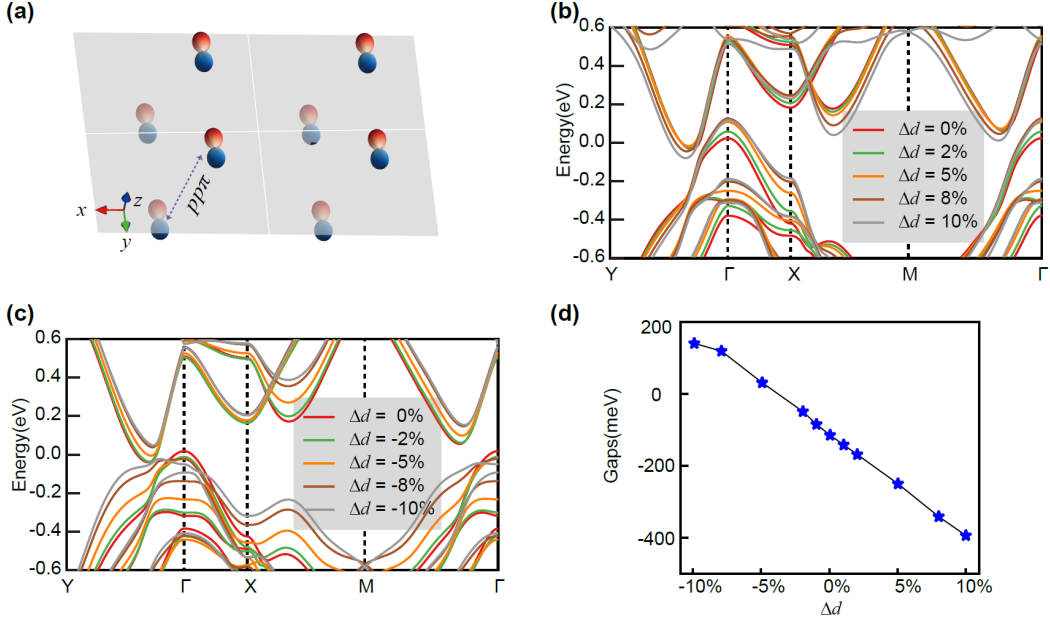

Fig. S2. Orbital analysis and first-principles calculations along $z$ direction. (a) Schematic plot of wave functions of $p_y$ orbitals. Four unit cells are plotted and the $pp\pi$-like bond along $z$ direction is indicated by double sided arrows. In-plane lattice constants $a$ and $b$ are along $x$ and $y$ directions, respectively. (b) and (c) Band structures with positive and negative $\Delta d$, respectively, calculated by the HSE03 hybrid functional. (d) The calculated relation between $\Delta d$ and gap size.

Since there is only one W atom layer, the related $d_{xz}$ orbital (see Fig. 1(b)) has no explicit bonds along $z$ direction, thus the $p_y$ orbital formed $pp\pi$-like bonds dominate the strain reactions, as shown in Fig. S2(a). When increasing/decreasing the atomic layer distance $\Delta d$, the $pp\pi$-like bonds get weaker/stronger, which will cause an energy increase/decrease of valence band upon a tensile/compressive strain. Thus a compressive strain along $z$ direction can also reduce the band overlap and even open a gap.

Here we tune atomic layer interval $d$ (marked in Fig. 1(a)) to simulate the influences of strains along $z$ direction. As shown in Fig. S2(b) and S2(c), when enlarging $\Delta d$ with a positive value, the CB (VB) will drop (rise) in energy and as a result the overlap between CB and VB is enhanced. On the contrary, decreasing Δ$d$ to a negative value, the CB will increase in energy and the VB will decrease in energy, which reduces the band overlap. Finally, the relation between Δ$d$ and gap value is shown in Fig. S2(d), which indicates an insulating gap will open under a proper compressive strain along $z$ direction.

## 2. Experimental evidences of strain effects along z direction

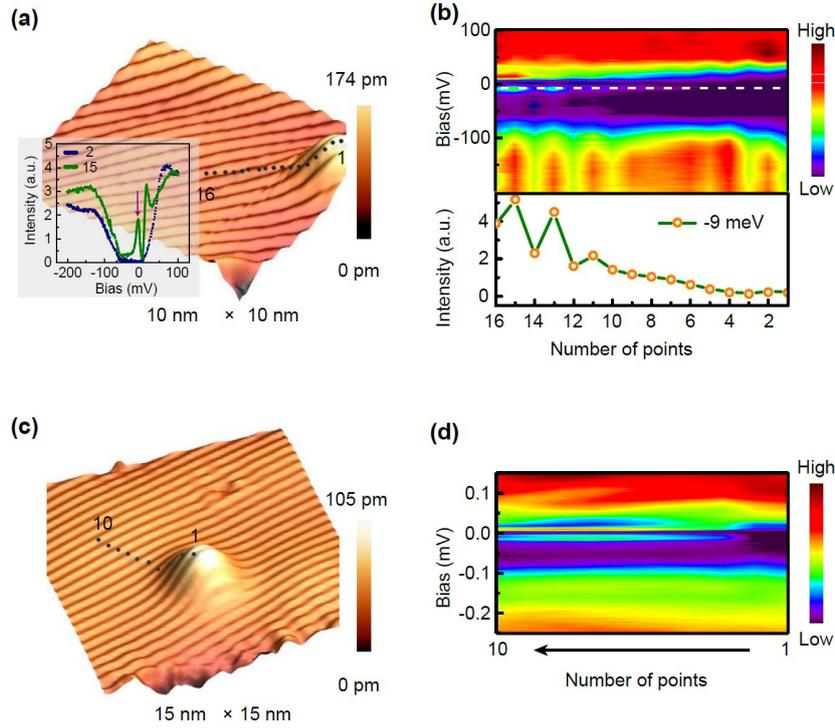

Fig. S3. Effects of bumps on band structures. (a) A Bump on WTe$_2$ films. 16 dI/dV spectra data points are taken in the different locations marked by black dots. Two typical dI/dV spectra taken on the bump (number 2) and away from the bump (number 15) are shown in the inset. (b) Spatially resolved color image of 16 dI/dV spectra data. Spatial distribution of intensity of in-gap states (-9meV) along the white dotted line is extracted and shown at the bottom. (c) Another bump on WTe$_2$ films. 10 dI/dV spectra data points are taken in the different locations marked by black dots. (d) Spatially resolved color image of 10 dI/dV spectra.

Although *z*-strains are difficult to control in experiments compared with in-plane strains, it is useful for band engineering according to both orbital analysis and first-principles calculations. However, due to the limitation of STM, we can only give a qualitative analysis of *z*-strains by measuring some bumps, as shown in Fig. S3(a) and S3(b). The dI/dV spectra taken away from the bump show semimetallic characters, while an insulating gap ~ 55 meV opens on top of the bump (see inset of Fig. S3(a)). A series of dI/dV spectra are taken from the bump top to the flat ground, as shown in Fig. S3(b). A full gap exists in the vicinity of the bump, and the gap size decreases when gradually moving away from the bump. Finally, the gap closes in the flat region. This can be also clearly demonstrated by the spatial distribution of LDOS of in-gap states, as shown in bottom panel of Fig. S3(b), where the LDOS of in-gap states at -9 meV decreases to zero near the bump. Besides, an interference of electronic states is observed, especially for the states lower than -100 mV, which is also observed in 1T'-WSe$_2$ [1]. The data for another bump also show suppression to in-gap states (see Fig. S3(c) and S3(d)), indicating that the bump can uniformly reduce the band overlap. Based on the crystal structure, this gap opening is presumably due to the compress of interlayer distance (negative $\Delta d$) by bumps, as mentioned in orbital analysis part.

## III. Confirmation of in-plane strain effects by first-principles calculations

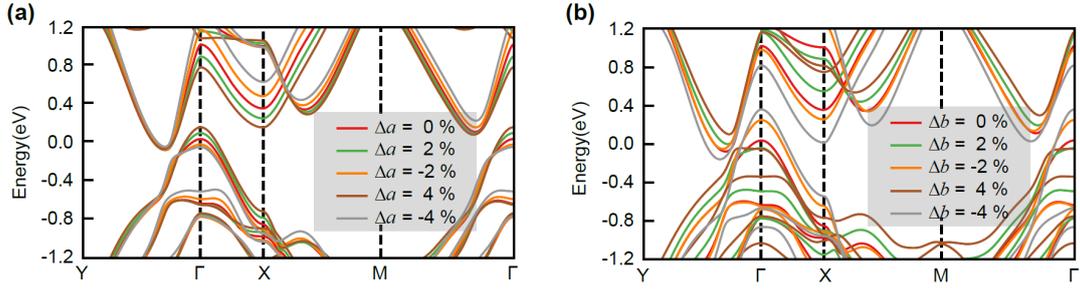

Fig. S4. Band structures of monolayer 1T'-WTe$_2$ by first-principles calculations. (A) and (B) Band structures under different uniaxial strains along *a* and *b* directions, respectively. Calculated by first-principles calculations.

Band structures calculated by using the HSE03 hybrid functional are shown in Fig. S4(a) and S4(b). It shows that the CB at Γ point will decrease/increase in energy under the tensile/compressive strain along *a* direction while opposite energy shifts would happen for VB in the same time. Along *b* direction, a tensile/compressive strain will result in the energy increase/decrease of CB. Meanwhile, the opposite energy shifts will happen for VB. These results by first-principles calculations are well matched with orbital analysis and further confirm that the gap size can be effectively tuned by uniaxial strains.

## IV. Gap as a function of in-plane strains calculated by different exchange-correlation functional

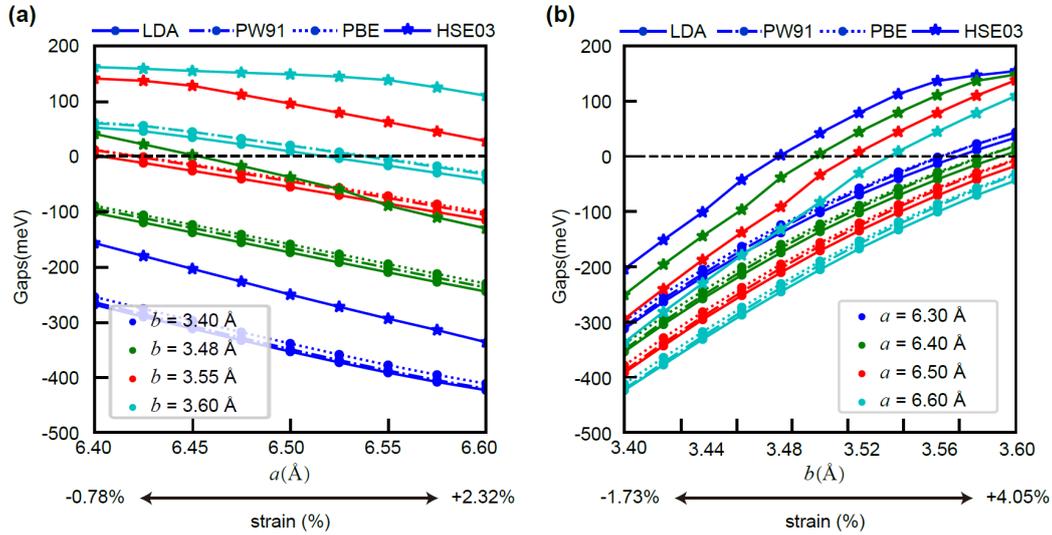

Fig. S5. Gap evolutions under uniaxial strains along $a$ and $b$ directions. (a) Gap values as a function of lattice constant $a$. The lattice constant $b$ is fixed at 0.340, 0.348, 0.355, 0.360 nm, which are represented by blue, green, red, and cyan lines, respectively. (b) Gap value as a function of lattice constant $b$ with lattice constant $a$ are fixed at 0.630, 0.640, 0.650, 0.660 nm, which are represented by blue, green, red, and cyan lines, respectively. The converted strain values based on lattice constants in flat regions are also shown at the bottom. The dot, dashed dot, solid lines with circle points, and solid lines with star points represent the PBE, PW91, LDA, and HSE03 potential functional, respectively.

For the above potential functional we used, gap values have the same behaviors: decrease under tensile strains along $a$ direction and increase under tensile strains along $b$ direction. For the given lattice constants, HSE03 gives the biggest gap value and LDA gives the smallest gap value. We choose the HSE03 hybrid functional because the gap size calculated by it shows the best match with our experimental results.

## V.  dI/dV spectra in distorted regions

### 1. dI/dV spectra in distorted regions in large scale

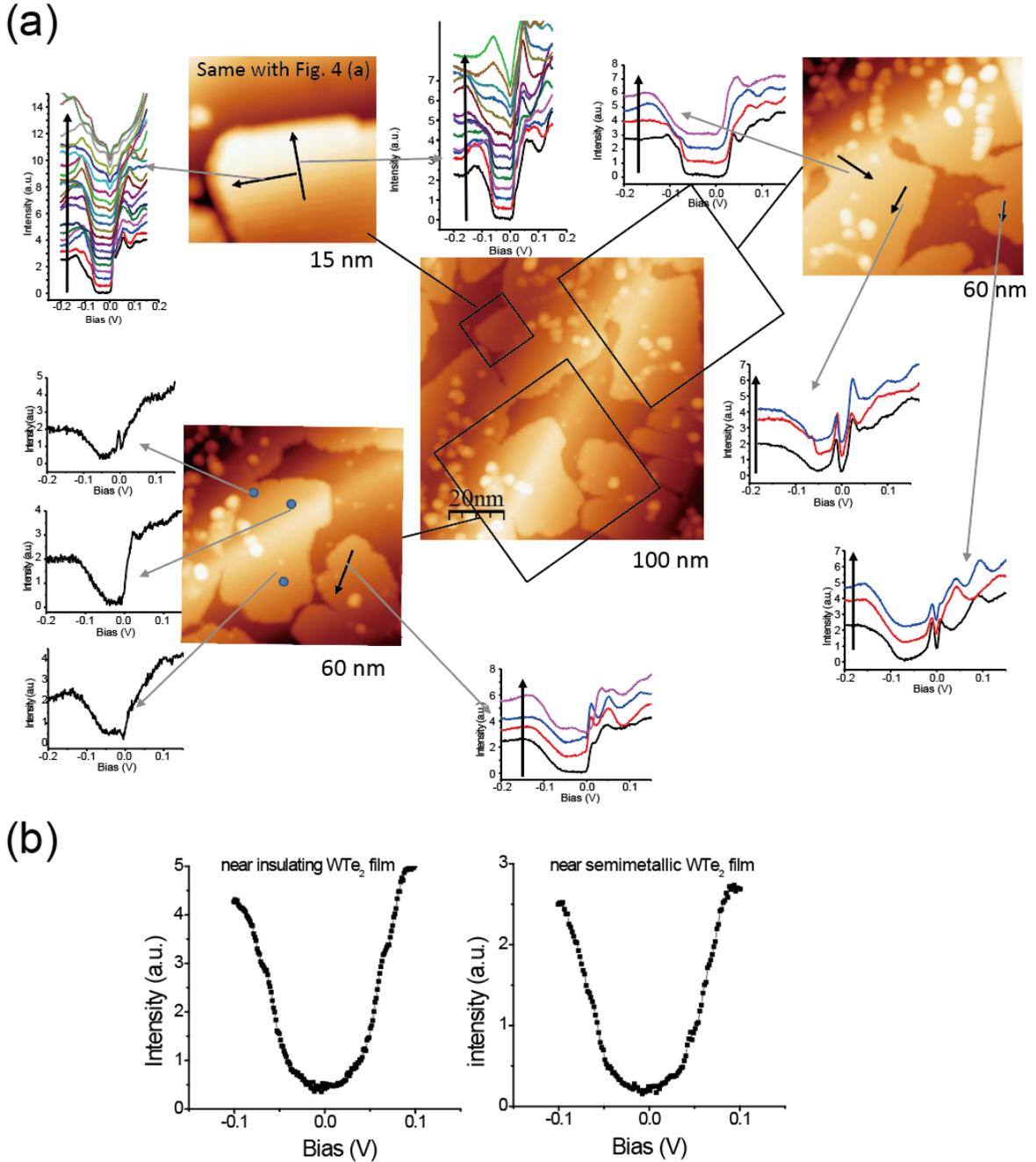

Fig. S6. Spatial dependency of dI/dV spectra in distorted regions. (a) Spatial-dependent spectra taken at different locations in a distorted region (100 nm × 100 nm, same region with Fig. 2(b)). (b) Comparison of electronic states of graphene substrate near insulating and semimetallic $WTe_2$ films.

The spectra taken at flat regions are uniform semimetallic type, unless the edge states emerge, as shown in Fig. 2(c) (also in Fig. S10 below), where the V-shape coulombs gap and a dip below Fermi level always exist unless reaching to film edges, no matter the size of the $WTe_2$ island. Both the dip and the coulomb gap do not reach to zero LDOS. While the spectra taken at different areas of distorted region are distinct, as shown in Fig. 2(d)-(f) and Fig. S6 (a), where the spectra show a remarkable spatial-dependency. An unique insulating type of spectra can be observed at some areas in distorted regions. And the gap size also varies in spatial. These results indicate that the insulating gap is probably related to spatially-dependent distortions.

To exclude the effects of different contributions form graphene substrate in distorted regions and flat regions, we compare the dI/dV spectra of graphene near an insulating $WTe_2$ island with that near a semimetallic $WTe_2$ island, as shown in Fig. S6(b). It is clear that the electronic states of graphene at these two regions are almost the same, especially at the energy scale [-100 meV, 0 meV], where the insulating gaps of $WTe_2$ emerge. We did not observed the strain induced Landau level in deformed graphene because this phenomenon requires vary local and strong strain conditions, such as the small nanobubbles observed by M. F. Crommie [2].

## 2. dI/dV spectra in distorted regions in small scale

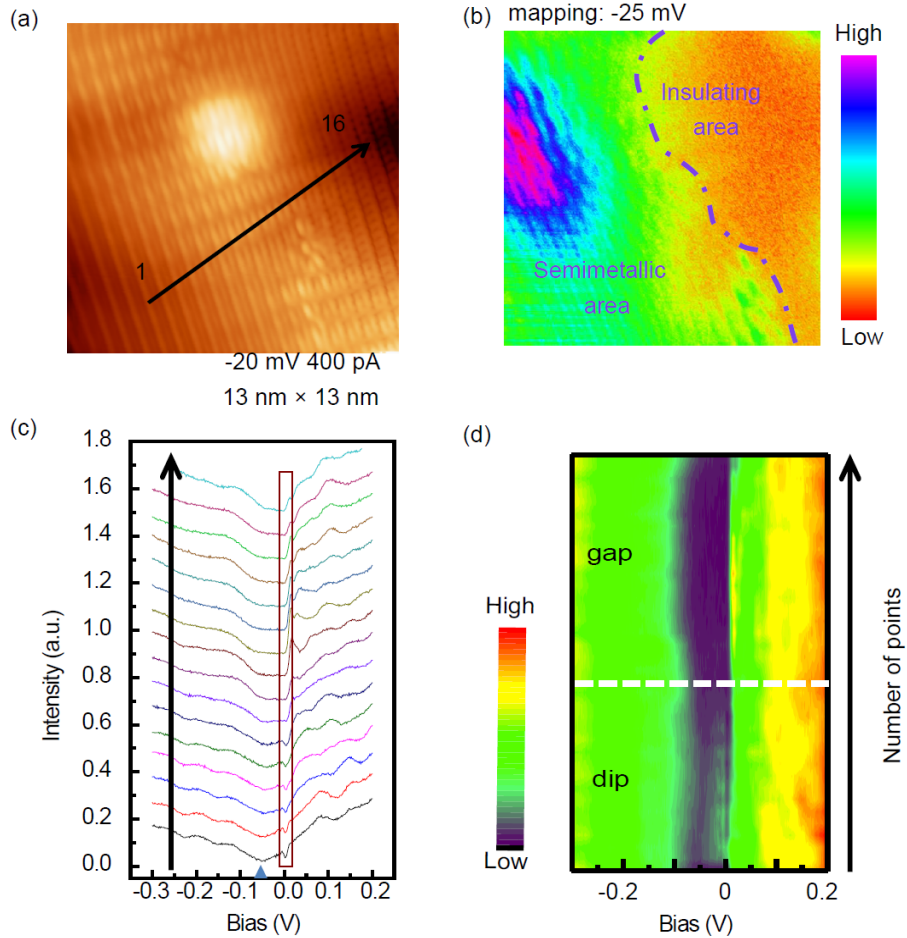

Fig. S7. Spatial dependenct-dI/dV spectra in a small distorted area. (a) Topography of a small distorted area (13 nm × 13 nm). (b) dI/dV mapping of the same area with the bias of -25 meV. (c) 16 dI/dV spectra taken along black arrow in (a). (d) Spatially-resolved color image of (c).

The dI/dV spectra results of small areas give a clearer view of the strain-dependent insulating gap, such as shown in Fig. S7(a), where the lattices show clear differences in right and left parts. The semimetallic type of spectra can be observed at the left part, while the insulting spectra exist at the right part. From Fig. S7(c), we can see that the dip at about -60 meV gradually evolves into large insulating gaps, while the coulomb gap residing at Fermi level is nearly unchanged, as long as the LDOS at around Fermi level is non-zero. Figure S7 (d) shows this spatially-dependent variation perfectly. The mapping using the bias inside the insulating gap (-25 meV) gives a phase boundary between semimetal and insulator (see Fig. S7 (b)).

# VI. Tuning Fermi level by electron doping

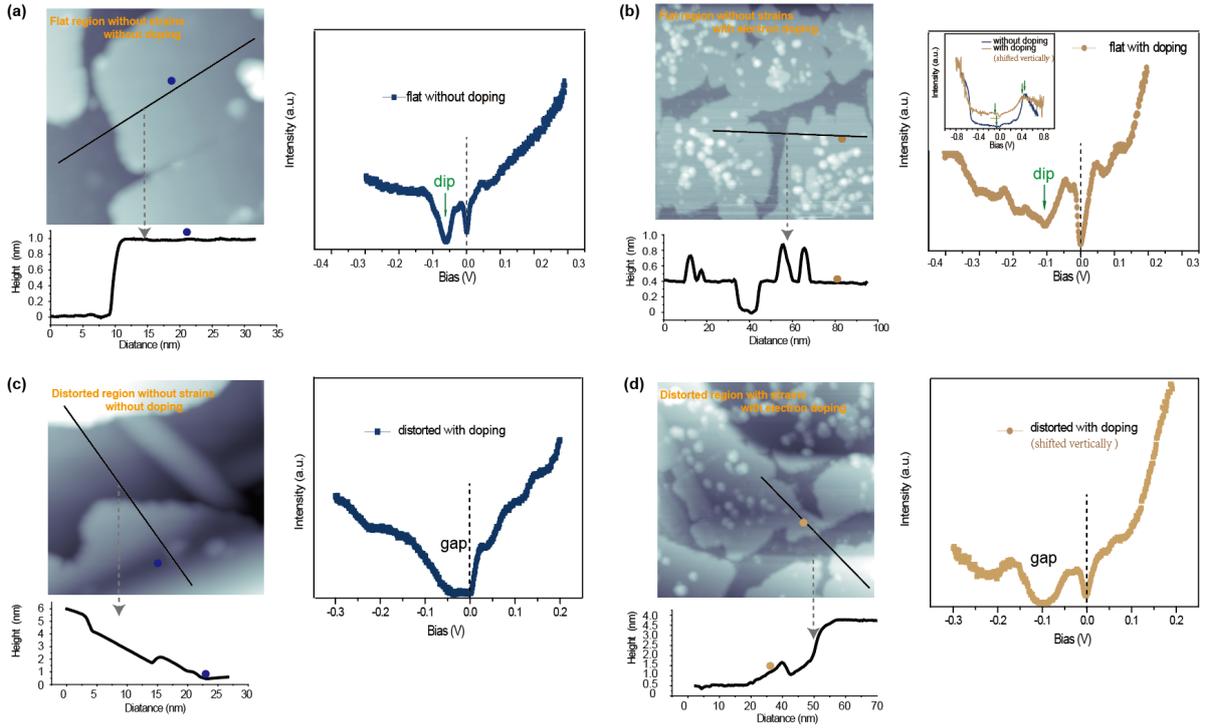

Fig. S8. Tuning Fermi level by electron doping. (a) and (b) Topography and typical spectra of WTe$_2$ films in flat regions with/without K doping. (c) and (d) Topography and typical spectra of WTe$_2$ films in distorted regions with/without K doping. The line profiles along black lines are shown at the bottom of topographic images.

To distinguish the strain induced U-shape band gap and coulomb gap and rule out the influences of coulomb interactions, we deposited some potassium (K) on the sample surface in order to dope electrons into WTe$_2$. Before depositing K atoms, the sample was firstly cooled down to 4.2 Kelvin to ensure that the K atoms deposited on WTe$_2$ but not flew away. The measurements were applied on WTe$_2$ sample after 2 min's deposition of K atoms.

Figure S8 (a) and S8 (b) shows the WTe$_2$ film in flat regions before/after K-doping, respectively. And the tunneling spectra taken on it show a dip feature below Fermi level as well as a coulomb gap residing at zero bias. Compared with the original spectra, the spectrum after electron-doping has an overall downward shift ~40 meV, and the dip now locates at around -100 meV. This shift can also be confirmed according to spectra in large energy scales (the inset of Fig. S8 (b)). Our results of electron-doping are similar with the previous research [3]. In distorted regions, fully band gaps can be acquired in samples both with (Fig. S8 (c)) and without (Fig. S8 (d)) electron-doping. It is worth noting that the band gap after electron-doping is far away from the Fermi level and is well separated with the coulomb gap which is still located at zero bias. These indicate that the band gaps opened by strain are irrelevant to Fermi level and are not caused by coulomb interaction.

# VII. Statistical process to extract the uniaxial relation between gap and lattice *a* and *b*

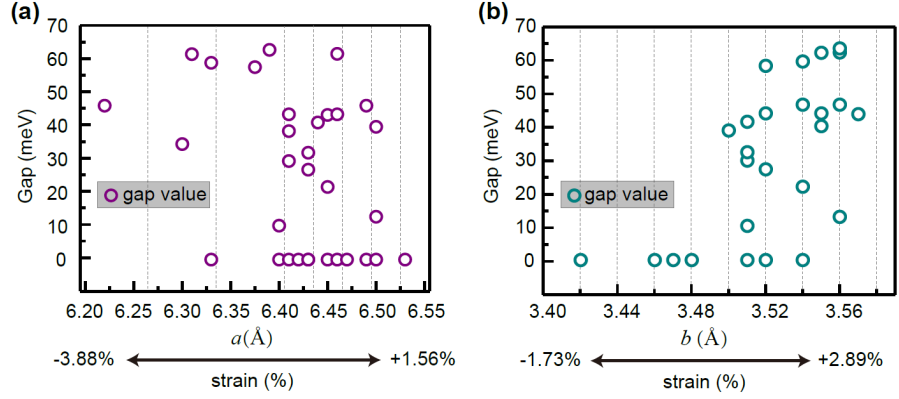

Fig. S9. Projected gap size along *a* and *b* axes. (a) and (b) the relations between gap values (raw data) and in-plane lattices *a* and *b*, respectively. The gray vertical dashed lines indicate the lattice intervals divided for averaging process. It should be noted that some of the data points may be covered because of the same parameters.

We have taken dozens of sets of data points, each of which contains a gap value and the corresponding in-plane lattices (*a*, *b*). In order to get the relations between gap value and uniaxial in-plane strain, we firstly project the same data according to their *a* and *b* lattice constants, respectively, as shown in Fig. S9 (a) and S9 (b). Since the gap value is determined by *a*, *b* and *d* jointly, an average process is required to eliminate influences of extra axes (in details: in relation between gap size and lattice constant *a*, the extra axes are *b* and *d*; in relation between gap size and lattice constant *b*, the extra axes are *a* and *d*). Then we divide the lattice *a* and *b* with step sizes of 0.03 Å and 0.02 Å, respectively, indicated by gray vertical dashed lines (the step sizes are 0.07 Å when *a* is smaller than 6.41 Å, because of the sparse distribution of data points). The average gap sizes in each interval are calculated and marked at the midpoints, as shown in Fig. 3 (b) and (c). After statistical process, the relations between gap size and uniaxial lattice a or b are clearer, because that the average process reduces the experimental error and, more importantly, the influences of distortions along extra directions.

And the influences of extra directions are transferred into the error bars, which are calculated by formula: $\sqrt{\frac{\sum_i(\Delta_i-\overline{\Delta})^2}{n}}$ , where $n$ is the number of data points in each interval, $\overline{\Delta}$ is the averaged gap value within each interval and the summation is made for all data points within each interval.

## VIII. Topological edges states on semimetallic WTe₂ islands

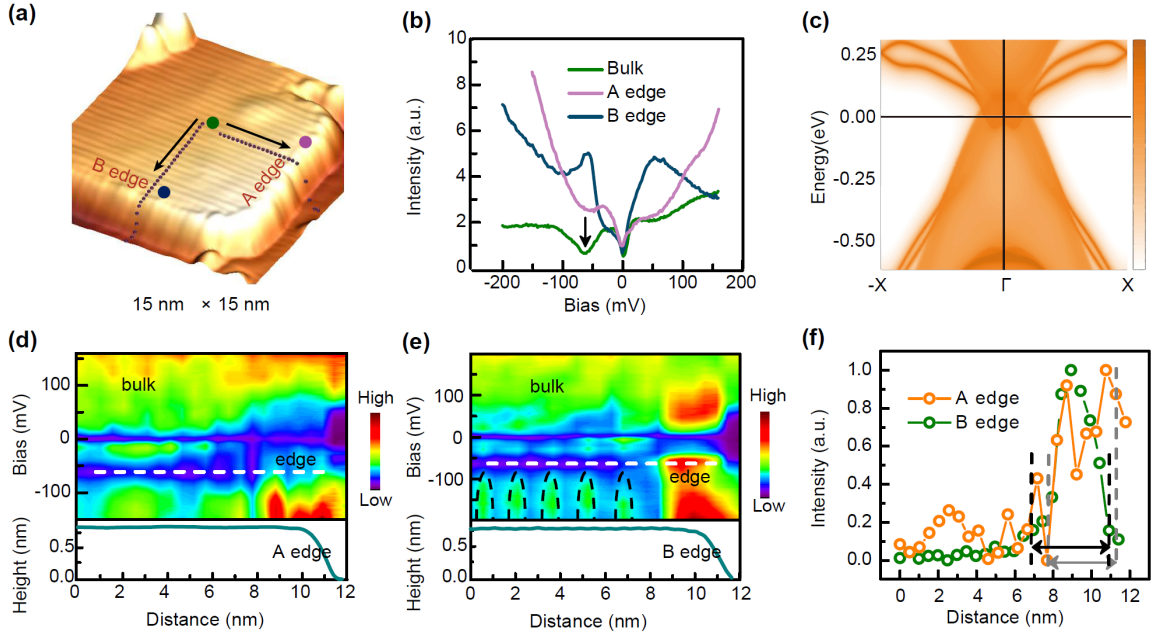

Fig. S10. Topological edge states of WTe₂ islands in flat regions. (a) Small island of 1T'-WTe₂ in flat region. (b) Typical dI/dV spectra taken in the interior and at two edges of the island. The corresponding locations are marked out by points with different colors in (a). (c) Calculated edge states along $A$ edge of semimetallic WTe₂ island. (d) and (e) Spatially resolved color images of a series of dI/dV spectra taken along dotted lines crossing $A$ edge and $B$ edge, respectively. (f) Spatial distribution of intensity of edge states (-64 mV) at two edges.

The edge states on semimetallic islands are similar to previous STM studies [4]. Figure S10(a) shows a small island of WTe₂ in flat region. In the bulk of the island, a semimetallic dI/dV spectrum with the intensity minimum of LDOS at ~ -60mV is acquired (green curve in Fig. S10(b)). Spectra taken at A-edge and B-edge are also shown in Fig. S10(b), both of which show remarkable increases of intensity around the Fermi level, indicating the existence of edge states. All three spectra show coulomb gaps at zero bias. The calculated topological edge states along $A$ edge are shown in figure S10(c), confirming the edge states induced enhancement of intensity around Fermi level. The calculated edge states along $B$ edge are shown in Fig. S12(b).

To further investigate the penetration of edge states, two series of dI/dV spectra crossing $A$ and $B$ edges are taken along dotted lines in Fig. S10(a), and their color images are shown in figure S10(d) and S10(e), respectively. It should be noted that the last two points of both lines are locate on the graphene substrate, where the spectra are totally different from that on WTe$_2$. The dip feature at ~ -64 mV is uniform in the bulk. When approaching to edges, the dip vanishes gradually and the edge states emerge. The spatial distributions of state intensity at -64 meV are extracted to characterize the penetration of edge states along *A* and *B* edges, respectively (see orange line and green line in Fig. S10(f)), where the penetration lengths are estimated to be ~ 3.6 nm for *A* edge and ~ 4.0 nm for *B* edge. The signals of standing waves are also observed in figure S10(e) (marked out by black arcs).

## IX. Edge states on other insulating WTe$_2$ islands

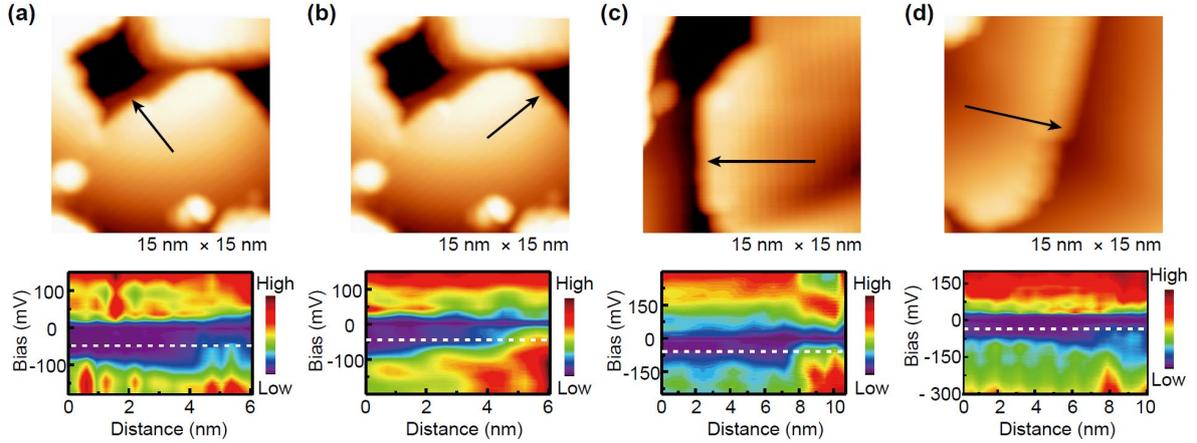

Fig. S11. Edge states of insulating WTe$_2$ islands. (a)-(d) Spatially resolved color images of dI/dV spectra taken crossing different edges of insulating WTe$_2$ islands along the black arrows.

Here we show edges states on several insulating islands. The topography and color images of dI/dV spectra crossing island edges are shown in Fig. S11. Although the gap sizes are different and not uniform on each island, signatures of edge states can always be observed. The penetration lengths are all in the range from 2 to 4 nm. These results demonstrate the robustness of topological edge states under various experimental strain conditions, which is consistent with the calculation results.

## X. Calculated edge states along B edge

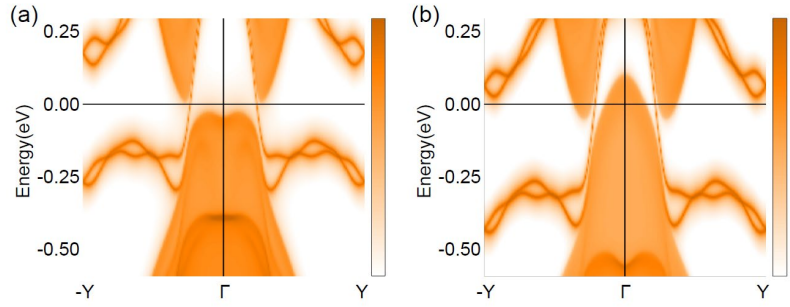

Fig. S12. Calculated edge states along *B* edge. (a) Calculated *B* edge states of insulating WTe$_2$ island. (b) Calculated *B* edge states of semimetallic WTe$_2$ island.

Calculated edges states along *B* edge are shown in Fig. S12. Using the lattice constants of the insulating island shown in Fig. 4(b), the calculated bands along *B* edge also show the topologically protected edge states, as shown in Fig. S12(a). Similar calculations are applied for the semimetallic island shown in Fig. S10(a), and the results are shown in Fig. S12(b).

## XI. Methods

### 1. Treatments for the substrate

Bilayer graphene created from 6H-SiC (0001) is selected as the substrate to grow monolayer 1T'-WTe$_2$. The 6H-SiC is first heated to 600 °C to degas for 12 hours in the MBE chamber with a base pressure of $2\times10^{-10}$ Torr. It is then heated to 900 °C for 30 mins and then to 1300 °C for 10 mins. Finally, it is gradually cooled to room temperature. By this process we obtain a bilayer graphene surface on the 6H-SiC substrate. If the maximum temperature during the heat treatment exceeds 1300 °C, there are likely to be some surface deformations, such as irregular bends and deep holes.

### 2. MBE growth of 1T'-WTe$_2$

The monolayer 1T'-WTe$_2$ films are grown on bilayer graphene by the MBE method in a vacuum chamber with base pressure of $2\times10^{-10}$ Torr. The tellurium (Te) source is a knudson diffusion cell and the tungsten (W) source is an electron beam evaporator. The Te source material has a purity of 99.999% and the W has a purity of 99.995%. The flux ratio between W and Te is about 1 : 20. The growth process is monitored by reflection high-energy electron diffraction (RHEED). After growth the sample is annealed at 260 °C in a Te environment for 2 hours and then without the Te environment for 30 mins to enhance the film quality.

### 3. STM measurement

In situ STM measurements are carried out after sample growth. The STM chamber has a base pressure of $1\times10^{-10}$ Torr. The sample is cooled to 4.2 K before all measurements. The topographic images are obtained using a constant current mode. Before gathering data, the STM tip is first calibrated on a silver surface.

### 4. **First-principles calculations**

The calculations are performed in the framework of density functional theory as implemented in the VASP [5] package using the projector augmented wave (PAW) method[6]. The Brillouin zone is sampled in 7 × 14 × 1 Monkhorst-Pack *k*-point grid, and the energy cutoff of the plane wave basis is 400 eV. We compared the calculations from different exchange-correlation functionals including GGA (Perdew-Wang 91 (PW91), Perdew-Burke-Ernzerhof (PBE)) and LDA, and the results are almost identical. We also used the HSE03 hybrid functional to improve the band gap calculations [7]. *Ab-initio* based tight-binding calculations were performed using the WANNIER90 interface [8]. The edge states were calculated by the iterative Green's function approach [9].